\documentclass{jetpl}
\twocolumn

\usepackage[cp1251]{inputenc}
\usepackage[english,russian]{babel}

\lat

\usepackage{graphicx}
\usepackage{color}
\usepackage[pdfborder={0 0 0}, colorlinks=true, linkcolor=blue, citecolor=blue, urlcolor=blue]{hyperref}

\title{Band structure of tungsten oxide W$_{20}$O$_{58}$ with ideal octahedra}

\rtitle{Band structure of tungsten oxide W$_{20}$O$_{58}$ with ideal octahedra}

\sodtitle{Band structure of tungsten oxide W$_{20}$O$_{58}$ with ideal octahedra}

\author{M.\,M.\,Korshunov\/$^{1,}$\thanks{mkor@iph.krasn.ru}, I.\,A.\,Nekrasov\/$^{2}$, N.\,S.\,Pavlov\/$^2$, A.\,A.\,Slobodchikov\/$^{2,}$\thanks{stalfear@gmail.com}}

\rauthor{M.\,M.\,Korshunov, I.\,A.\,Nekrasov,  N.\,S.\,Pavlov, A.\,A.\,Slobodchikov}

\sodauthor{Korshunov, Nekrasov, Pavlov, Slobodchikov}

\address{$^{1}$Kirensky Institute of Physics, Federal Research Center KSC SB RAS, Akademgorodok, 660036 Krasnoyarsk, Russia \\
$^{2}$ Institute of Electrophysics, Russian Academy of Sciences, Ural Branch, 620016 Ekaterinburg, Russia}

\abstract{The band structure, density of states, and the Fermi surface of a tungsten oxide WO$_{2.9}$ with idealized crystal structure (ideal octahedra WO$_6$ creating a ``square lattice'') is obtained within the density functional theory in the generalized gradient approximation. Because of the oxygen vacancies ordering this system is equivalent to the compound W$_{20}$O$_{58}$ (Magn\'{e}li phase), which has 78 atoms in unit cell. We show that 5$d$-orbitals of tungsten atoms located immediately around the voids in the zigzag chains of edge-sharing octahedra give the dominant contribution near the Fermi level. These particular tungsten atoms are responsible of a low-energy properties of the system.}


\begin{document}

\maketitle

\textbf{1. Introduction.}
Superconductivity, as one of the fundamental ground states in solid state physics, is realized sometimes in the most unexpected cases. These are both high-temperature superconducting cuprates~\cite{bednorz-muller}, which are dielectrics in the underdoped case, as well as pnictides and iron chalcogenides~\cite{y_kamihara_08,SadovskiiReview,HirschfeldKorshunov2011,Korshunov2014}, although under normal conditions iron is a ferromagnet. These systems are unusual superconductors, i.e. having anisotropic momentum dependence of the order parameter. They are related to tungsten oxides by the presence of a partially filled $d$-shell. Oxygen non-stoichiometric WO$_{3-x}$ tungsten trioxide compounds have been known for a long time and their structure and properties have been well studied~\cite{Bursill1972,Sahle1983}. However, more recently, the discovery of superconductivity in the compound WO$_{2.9}$ with $T_c=80$~K and with $T_c=94$~K when intercalated with lithium~\cite{Shengelaya2020} has been reported. This was preceded by the observation of superconductivity near the domain walls in WO$_{3-x}$\cite{Aird1998}, thin films~\cite{Kopelevich2015} and in WO$_3$ with surface deposited sodium, Na$_{0. 05}$WO$_3$~\cite{Reich1999}, which led to the prediction of the possibility of superconductivity realization in WO$_{3-x}$~\cite{Shengelaya2019}.

\begin{figure*}[h]
	\centering
	\includegraphics[width=0.9\linewidth]{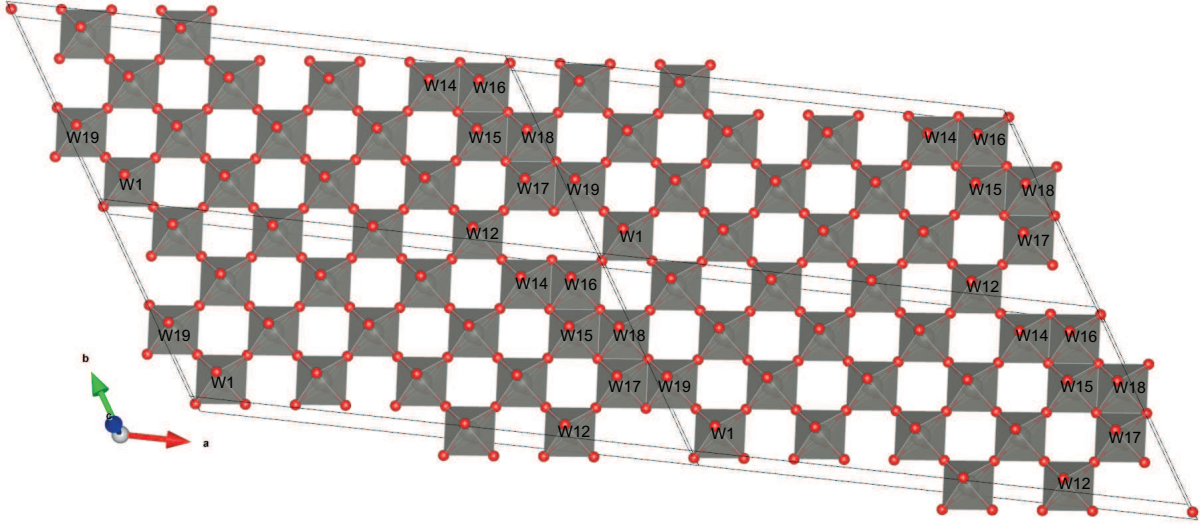}
	\caption{Fig. 1. Idealized crystal structure of the W$_{20}$O$_{58}$ supercell ($2 \times 2 \times 1$).}
	\label{fig:structure}
\end{figure*}

Despite the long history of research on tungsten oxides as of today there are only a few works on the band structure calculations of WO$_3$. These are works on the electronic structure of bulk samples, thin films and clusters~\cite{Hamdi2016,Wijs1999,Gonzalez-Borrero2010,Johansson_2013,Ping2013,Tsipis2000,Stachiotti1997, Cora1996}, the role of oxygen vacancies~\cite{Migas2010_1,Wang2011_wo3_jpc,Wang2011_wo3_prb,KARAZHANOV2003,Mehmood2016} and cationic doping~\cite{Walkingshaw2004,Tosoni2014,Hjelm1996,Ingham2005,Huda2009,Cora1997,Huda2008,Chen_2013}.
Calculations for Magn\'{e}li phases with ordered oxygen vacancies, WO$_{3-x}$, are described in only one work~\cite{Migas2010_2}, which illustrate that the compounds W$_{32}$O$_{84}$, W$_3$O$_8$, W$_{18}$O$_{49}$, W$_{17}$O$_{47}$, W$_5$O$_{14}$, W$_{20}$O$_{58}$ and W$_{25}$O$_{73}$ show metallic properties.

The ordering of oxygen vacancies in the system WO$_{3-x}$ leads to the appearance of quite large unit cells, which significantly complicates its description. And superconductivity is realized in the system W$_{20}$O$_{58}$ containing 78 atoms in a unit cell. Tungsten atoms coordinated by oxygen atoms form octahedra which are either corner-sharing or edge-sharing. The octahedra themselves are distorted, and the W-O-W bonds between the octahedroa are distorted as well, which leads to an additional complication of the description of the electronic structure of W$_{20}$O$_{58}$.

Since the foundation for building a superconducting state theory is the band structure of the normal phase, the calculation of the last one from the first principles will be the first step on this path. In this paper we obtained the band structure, density of states and Fermi surface for the compound WO$_{3-x}$ with ideal octahedra creating a ``square lattice'', which is the first approximation in the description of this complex compound.

\textbf{2. Structure and calculation results.}
W$_{20}$O$_{58}$ belongs to the family of oxides with the Magn\'{e}li structure and the general formula W$_n $O$_{3n-2}$~\cite{Bursill1972}. Space group is $P2$/$m$:$b$, lattice parameters are $a = 12.1$ \AA, $b = 3.78$ \AA, $c = 23.39$ \AA, $\beta = 95^{\circ}$~\cite{Magneli1949}. The crystal structure consists of WO$_6$ octahedra which are either corner-sharing or edge-sharing in the (100) plane.

As stated earlier in the compound W$_{20}$O$_{58}$ the WO$_6$ octahedra are distorted and the O-O bond length ranges from 2.63 to 2.72~\AA. In order to model the idealized crystal structure, all octahedra were made ideal with average oxygen-oxygen distance equal to 2.68~\AA. In this case, the bases of all ideal octahedra form a ``square lattice''. Fig.~\ref{fig:structure} shows the supercell $2 \times 2 \times 1$ for the idealized crystal structure W$_{20}$O$_{58}$.

To calculate the band structure, the density of states (DOS) and the Fermi surface we use the density functional theory (DFT) with all-electron full-potential linearized augmented-plane wave method (FP-LAPW) implemented via the Elk code~\cite{elk} together with the generalized gradient approximation (GGA)~\cite{jperdew96}. For the self-consistent ground state calculation we used a $8 \times 8 \times 8$ \textbf{k}-points grid in an irreducible Brillouin zone making sure that the results are almost indistinguishable from those for a $6 \times 6 \times 6$ grid.

\begin{figure}[h]
\begin{center}
 \includegraphics[width=0.7\linewidth]{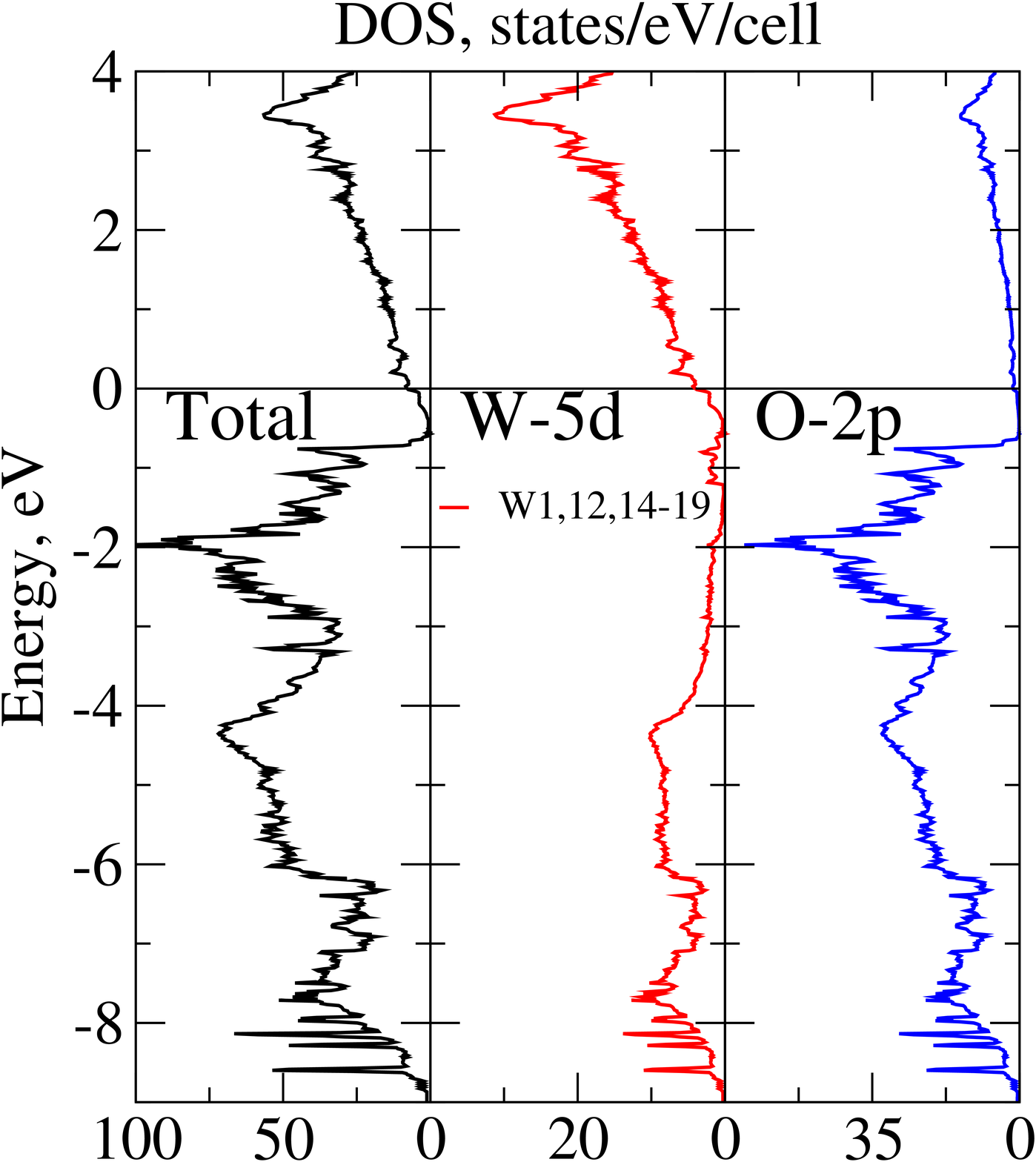}
 \caption{Fig. 2. Total DOS for W$_{20}$O$_{58}$ with idealized crystal structure (left), DOS for tungsten atoms W1,12,14-19 (center), DOS for oxygen atoms (right) in a wide energy range. Zero corresponds to the Fermi level.}
 \label{fig:dos_wide_bands}
\end{center}
\end{figure}

The density of states in the wide energy range are shown in Fig.~\ref{fig:dos_wide_bands}.
The top of the valence band from $-0.8$~eV to $-4.0$~eV is formed mainly by O-2$p$ states. In the region from $-4.0$~eV to $-9.0$~eV we see strong hybridization of W-5$d$ and O-2$p$ states. 

In the stoichiometric compound WO$_3$ tungsten W$^{6+}$ has a $5d^0$ configuration, that is, an empty 5$d$-shell and, therefore, a completely filled O-2$p$ shell, being a band dielectric. The oxygen deficit in WO$_{3-x}$ leads to electron doping and a finite conductivity value~\cite{Sahle1983}. This is clearly seen in our calculation as well, where the W-5$d$ states are almost empty and form the conduction band. At the Fermi level we can see only a low-intensity tail coming from W-5$d$ states (see Fig.~\ref{fig:dos_wide_bands}), which are filled with electrons due to an oxygen deficit compared to the stoichiometric composition of WO$_3$.

\begin{figure}[h]
\begin{center}
 (a)\includegraphics[width=0.6\linewidth]{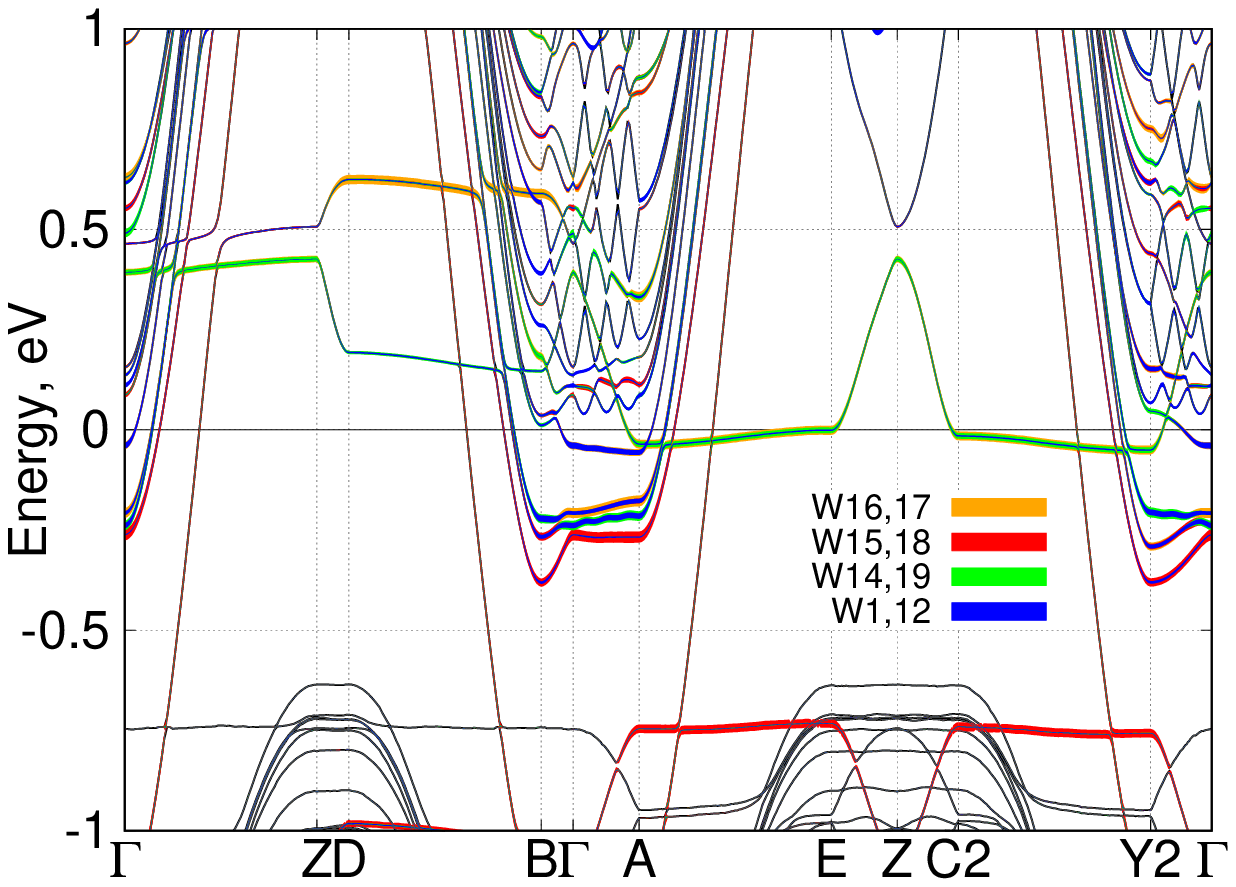}
 (b)\includegraphics[width=0.18\linewidth]{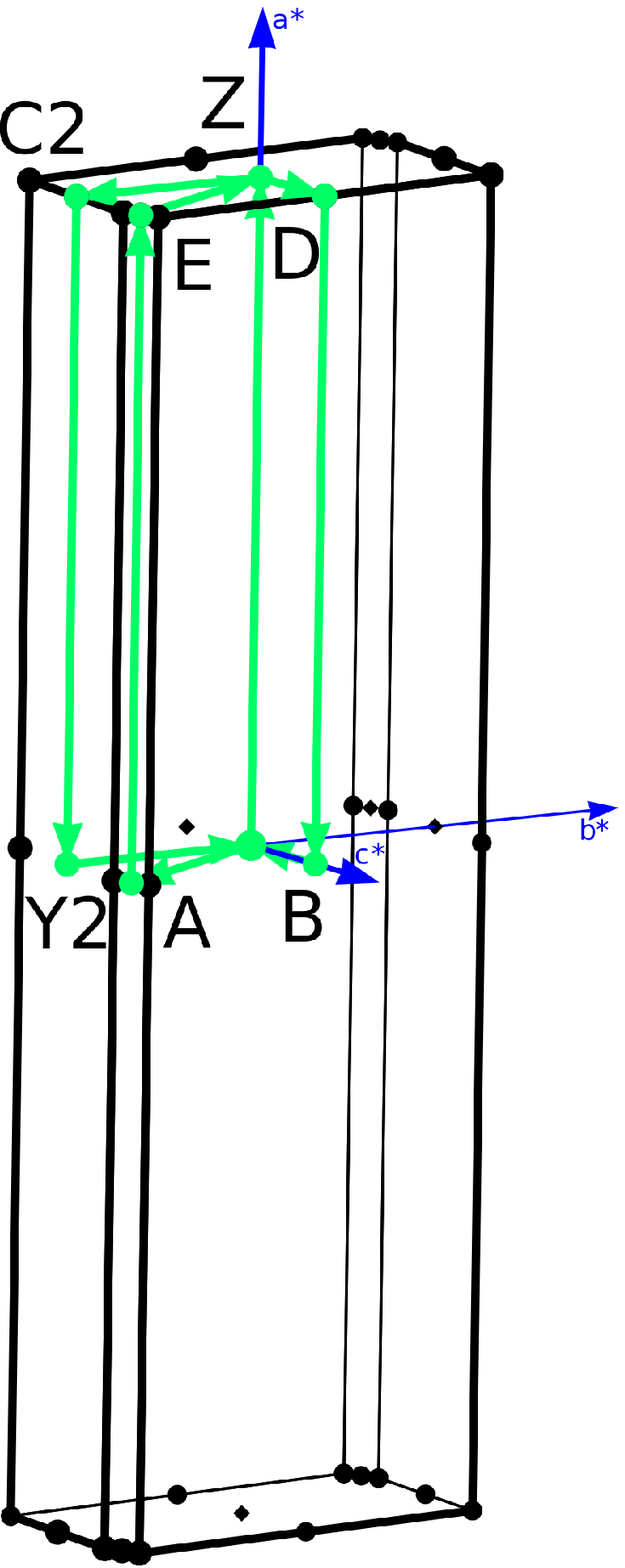}
 \caption{Fig. 3. (a) --- Band structure for W$_{20}$O$_{58}$ with idealized crystal srtucture near the Fermi level. Color denote the contribution of individual tungsten atoms W1,12,14-19. Zero corresponds to the Fermi level. (b) --- Brillouin zone for idealized crystal srtucture W$_{20}$O$_{58}$.}
 \label{fig:BZ_bands}
\end{center}
\end{figure}

We would like to emphasize the presence of flat bands at the Fermi level in the direction $A-E$ and near it, as well as in the direction $\Gamma-A$, which are shown in Fig.~\ref{fig:BZ_bands}(a). For the bands structure we used highly symmetric \textbf{k}-points and corresponding directions generated with the SeeK-path~\cite{Hinuma2017} tool. You can see them in Fig.~\ref{fig:BZ_bands}(b).

To demonstrate which states form these flat bands Fig.~\ref{fig:BZ_bands}(a) shows the bands structure with the contributions of individual atoms in the vicinity of the Fermi level. One can see that the flat bands are formed by 5$d$-states of W1,12,14-19 tungsten atoms, which are arranged around the voids in zigzag chains of edge-sharing octahedra (see Fig.~\ref{fig:structure}). Located directly around the voids atoms W14,19 and W16,17 give the greatest contribution. It is also worth saying that the 5$d$ states of W1,12,14-19 tungsten provide about 70\% to the value of the total state density at the Fermi level.
Note that some ``chaotic'' bands structure in the $\Gamma-A$ direction, visible above the Fermi level, is nothing but multiple crossing of bands, resulting from the rather small volume of the Brillouin zone and the large number of atoms in the unit cell, split between themselves by small hybridization interaction. 

\begin{figure}[h]
	\begin{center}
		\includegraphics[width=0.3\linewidth]{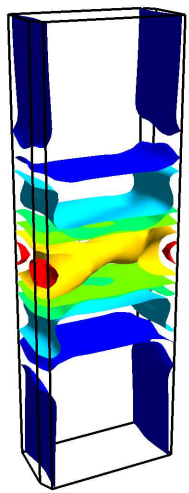}
		\caption{Fig. 4. Fermi surface for idealized crystal srtucture W$_{20}$O$_{58}$.}
		\label{fig:fermi_surfaces}
	\end{center}
\end{figure}

Fig.~\ref{fig:fermi_surfaces} shows the Fermi surface for W$_{20}$O$_{58}$ with an idealized crystal structure. The corresponding Fermi surface contains six sheets. The sheets located near the $\Gamma$-point (yellow and red) are clearly three-dimensional, while the other sheets are quasi-two-dimensional. Note that the flat bands at the Fermi level in the $A-E$ direction form rather large two-dimensional hole pockets at the corners of the Brillouin zone. 

\textbf{3. Conclusion.}
We have studied the compound W$_{20}$O$_{58}$ with an idealized crystal structure via the first-principles DFT-GGA calculation. Despite the large number of atoms (seventy eight) in the unit cell the main contribution to the states near the Fermi level originate from the 5$d$-orbitals of tungsten. These atoms are located directly around the voids in the zigzag pattern of edge-sharing octahedra. Thus, it is this zigzag pattern, disordering the ideal ``checkerboard'' arrangement of octahedra, are responsible for conductivity and other effects related to the states near the Fermi surface. On the one hand, we are dealing with a complex crystal structure with a huge unit cell, which is caused by the disordered arrangement of some tungsten atoms, and on the other hand, the 5$d$-orbitals of these atoms determine the low-energy physics of the compound WO$_{3-x}$.

We would like to thank S.G. Ovchinnikov and M.V. Sadovskii for useful discussions. 
This work was supported by RFBR and Government of Krasnoyarsk Territory and Krasnoyarsk Regional Fund of Science to the Research Projects ``Electronic correlation effects and multiorbital physics in iron-based materials and cuprates'' grant No. 19-42-240007 (MMK), by RFBR grants No. 18-02-00281, 20-02-00011 (IAN, NSP, AAS), by the President of Russia grant for young scientists No. MK-1683.2019.2 (NSP and AAS). The computations were performed at ``URAN'' supercomputer of the Institute of Mathematics and Mechanics of the RAS Ural Branch.


\vfill\eject

\end{document}